\documentclass{aa}
\usepackage{graphicx}
\usepackage{amsmath}
\usepackage{multirow}
\usepackage{placeins}
\usepackage{gensymb}
\usepackage{natbib,twoopt}

\usepackage{txfonts}
\bibpunct{(}{)}{;}{a}{}{,}    
\usepackage[colorlinks,citecolor=blue,linkcolor=blue,urlcolor=blue,bookmarks=false,hypertexnames=true]{hyperref}

\begin{document}

   \title{A variable magnetic disc wind in the black hole X-ray binary GRS 1915+105?}


   \author{A. Ratheesh\inst{1,2},
          F. Tombesi\inst{1,3,4,5,6},
          K. Fukumura\inst{7},
          P. Soffitta\inst{2},
          E. Costa\inst{2} \and 
          D. Kazanas\inst{6} 
          }

\institute{Department of Physics, Tor Vergata University of Rome, Via della Ricerca Scientifica 1, I-00133 Rome, Italy\\
              \email{ajay.ratheesh@roma2.infn.it}
         \and
             INAF - IAPS, Via Fosso del Cavaliere 100, I-00133 Rome, Italy
        \and INAF - Astronomical Observatory of Rome, Via Frascati 33, I-00078 Monte Porzio Catone (Rome), Italy 
        \and INFN - Tor Vergata, Via della Ricerca Scientifica 1, I-00133 Rome, Italy
        \and Department of Astronomy, University of Maryland, College Park, MD 20742, USA
        \and NASA/Goddard Space Flight Center, Greenbelt, MD 20771, USA
        \and James Madison University, 800 South Main Street, Harrisonburg, Virginia 22807, USA}

\date{Accepted XXX. Received YYY; in original form ZZZ}



 
  \abstract
   {
   GRS 1915+105 being one of the brightest transient black hole binary (BHB) in the X-rays, offers a unique test-bed for the study of the connection between accretion and ejection mechanisms in BHBs. In particular, this source can be used to study the accretion disc wind and how it depends on the state changes in BHBs.
   }
   {Our aim is to investigate the origin and geometry of the accretion disc wind in GRS 1915+105. This study will provide a basis for planning future observations with the X-ray Imaging Spectroscopy Mission (XRISM), and it may also provide important parameters for estimating the polarimetric signal with the upcoming Imaging X-ray Polarimetry Explorer (IXPE).
   }
   {We analysed the spectra of GRS 1915+105 in the soft $\phi$ and hard $\chi$ classes, using the high resolution spectroscopy offered by Chandra HETGS. In the soft state, we find a series of wind absorption lines that follow a non linear dependence of velocity width, velocity shift and equivalent width with respect to ionisation, indicating a multiple component or stratified outflow. In the hard state we find only a faint Fe XXVI absorption line. We model the absorption lines in both the states using a dedicated MHD wind model to investigate a magnetic origin of the wind and to probe the cause of variability in the observed lines flux between the two states. 
   }
   {The MHD disc wind model provides a good fit for both states, indicating the possibility of a magnetic origin of the wind. The multiple ionisation components of the wind are well characterised as a stratification of the same magnetic outflow. We find that the observed variability in the lines flux between soft and hard states cannot be explained by photo-ionisation alone but it is most likely due to a large (three orders of magnitude) increase in the wind density. We find the mass outflow rate of the wind to be comparable to the accretion rate, suggesting a intimate link between accretion and ejection processes that lead to state changes in BHBs.}
   {}


\titlerunning{A variable magnetic disc wind in GRS 1915+105}
\authorrunning{A, Ratheesh et al}
\maketitle

\section{Introduction}
Black hole binaries (BHBs) are the best candidates to study the connection between accretion and ejection mechanisms due to their high variability and high X-ray flux. The timescale of variability is comparable to human timescales and hence the whole cycle of accretion and ejection can be studied for a single system. Even though BHBs may be seen as a scaled down versions of active galactic nuclei (AGN), their observational signatures may partially differ due to the different source of accreting material, i.e. the companion star or galactic gas, respectively, and also the radiation field that is predominantly governed by X-rays. \\
\\
The variability of low mass X-ray binaries (LMXBs) can be classified using the hardness intensity diagram (HID) in X-rays as many of them exhibit a hysteresis loop within it \citep{2000A&A...355..271B}. Different points in the loop can be attributed to different accretion-ejection geometries. The presence of a powerful jet is generally seen when the source transits from hard to soft state  \citep{Fender_2004}, establishing the connection between BHB states and the launch of the jet. However GRS 1915+105, even though being a LMXB, does not exhibit a hysteresis loop in the HID, and was in a comparatively softer state, until recently when it became harder in spectra and fainter in brightness \citep{MillerAtel2019_2019ATel12743....1M}. Outflows in the form of accretion disc winds have also been seen in several LMXBs \citep{2002ApJ...567.1102L,2004ApJ...601..450M, 2006AN....327..997M, 2006ApJ...646..394M}, mainly but not necessarily in the soft state \citep{2009Natur.458..481N}. Despite its clear detection, there is no consensus yet on how and where the wind is launched from the accretion disc, and how it affects the state transitions and the launching of the jet. In GRS 1915+105, it was seen that during the states in which the wind was present, the jet was either weak or absent, and conversely \citep{2009Natur.458..481N}. Thus, understanding the launching mechanism of the wind is required to study the interplay between the wind, the jet and the accretion states of BHBs. Hence, an insight into the launching mechanism, and an estimate of the launching site of the wind, along with the wind mass flux can help in determining the unknown parameters of the state transitions in X-ray binaries.  \\
\\
The mechanisms responsible for the launching of disc winds is still debated. The main phenomena that are thought to be responsible for it are (i) thermally driven, (ii) radiation pressure driven, and (iii) magnetically driven. There has been evidence for all the three processes \citep{Begelman1983_1983ApJ...271...70B,Ueda2004_2004ApJ...609..325U,Neilsen2013_2013AdSpR..52..732N,Trigo2016_2016AN....337..368D,Neilsen2016_2016ApJ...822...20N,done2018_2018MNRAS.473..838D,2017NatAs...1E..62F,Miller2006_2006Natur.441..953M}, as well as two component winds where two different mechanisms co-exist, namely MHD and thermal \citep{Neilsen2012_2012ApJ...750...27N}. Thermally driven winds are launched when the X-rays near the compact object heat the disc surface to the Compton temperature ($T_C$) \citep{Begelman1983_1983ApJ...271...70B,Tombesi2010_2010A&A...521A..57T}. The wind is launched at a radius where the plasma thermal velocity is greater than the local escape velocity. However, thermal winds are also possible from 0.1 $R_{C}$ \citep{woods1996_1996ApJ...461..767W}, where $R_{C}$ is the Compton radius. Radiation pressure driven winds are due to the line pressure exerted on partially ionised elements \citep{Castor1975_1975ApJ...195..157C}. Disc winds can also be accelerated by the disc vertical magnetic pressure gradients or by the disc centrifugal forces in combination with magnetic fields \citep{Fukumura2010_2010ApJ...715..636F,2017NatAs...1E..62F,Blanford1982_1982MNRAS.199..883B,Contopoulos1994_1994ApJ...432..508C,Contopoulos1995_1995ApJ...450..616C,Ferreira1997_1997A&A...319..340F}. There are also evidences suggesting that disc winds in BHBs are preferentially seen with equatorial geometry \citep{2012MNRAS.422L..11P}.\\ \\ 
During the last decade significant improvement has been made in modelling accretion disc winds in BHBs using photo-ionisation codes. GRS 1915+105, GRO 1655-40, 4U 1630-47 and H 1743-322 are some of the BHBs which show the presence of disc winds as seen in terms of intense absorption lines in their X-ray spectra \citep{Miller2015_2015ApJ...814...87M,Miller2016_2016ApJ...821L...9M,Miller2006_2006Natur.441..953M,Miller2008_2008ApJ...680.1359M,Kallman2009_2009ApJ...701..865K,Neilsen2012_2012ApJ...750...27N,Ueda2009_2009ApJ...695..888U}. Several photo-ionisation modelling of these sources points towards a multiple component outflow or a MHD origin for the wind \citep{Miller2008_2008ApJ...680.1359M,Kallman2009_2009ApJ...701..865K,Miller2015_2015ApJ...814...87M,Miller2016_2016ApJ...821L...9M,Miller2006_2006Natur.441..953M}. However, there is a paucity of detailed MHD models that could explain the wind origin by directly fitting the data. \cite{2017NatAs...1E..62F} in the case of GRO J1655-40 being one of these examples.\\ 
\\
In the case of GRO J1655-40, \cite{Miller2006_2006Natur.441..953M} and  \cite{Kallman2009_2009ApJ...701..865K} using photo-ionisation modelling, argue that the wind cannot be explained with one component of absorption in the soft state. They estimated the launch radius from the ionisation parameter ($\xi$) and compared with the Compton radius ($R_{C}$), to disfavour the possibility of a single component thermal origin. \cite{Kallman2009_2009ApJ...701..865K} found that the lines with lower blueshift and larger launching radius from the black hole may be consistent with a thermal wind, while Fe K lines with a shorter launching radius and higher blue-shift favours a magnetic origin. \cite{Neilsen2012_2012ApJ...750...27N} made a comparison of the wind in the soft and hard states of GRO 1655-40, and found that photo-ionisation alone cannot explain the disappearance of many of the absorption lines in the hard state. \cite{Neilsen2012_2012ApJ...750...27N} further argue that the presence of a hybrid wind (thermal and MHD) in the hard state, evolving into a two component wind in the soft state, as suggested also by \cite{Kallman2009_2009ApJ...701..865K} and \cite{Miller2006_2006Natur.441..953M}. However \cite{2017NatAs...1E..62F} using a physically self-consistent photo-ionised MHD wind model found that the two components identified by previous works may be well described by a single MHD wind launched from a large range of radii from the accretion disc. \cite{Miller2015_2015ApJ...814...87M,Miller2016_2016ApJ...821L...9M} showed, from a purely photo-ionisation point of view, that at least two or three absorption components were required to model the lines in the case of all of the above mentioned sources. Another argument disfavouring a thermal origin is the consistency of the launch radius, estimated from re-emission of the wind and the ionisation parameter \citep{Trueba2019_2019ApJ...886..104T,Miller2015_2015ApJ...814...87M}.\\
\\
From the recent NICER observations of GRS 1915+105, \cite{Neilsen2018_2018ApJ...860L..19N} found that the accretion disc wind is persistent, with absorption line flux changing depending on the state and count-rate. In this work we model the absorption lines in GRS 1915+105 in the two representative soft and hard states, using the same MHD model as in the case of \cite{2017NatAs...1E..62F}. We also explore a similar approach as in \cite{Ueda2009_2009ApJ...695..888U}, dividing the soft state observation into two epochs where there was a increase in flux. In section 2 we describe the observations and data reductions. In section 3 we show the lightcurve and the spectra of both soft and hard states. In section 5 we show the phenomenological scaling relations and MHD wind modelling, and finally in section 6 we discuss the results with comparison to previous works. 
\section{Observations and Data reduction}
We considered the high resolution X-ray spectroscopy offered by Chandra High Energy Transmission Gratings Spectrometer (HETGS) data to study the wind variability in different states of GRS 1915+105. The source was observed by Chandra using the HETGS instrument for 22 times. However, we decided to analyse 2 observations, based on the presence and absence of a strong wind as reported in \cite{2009Natur.458..481N}. Since GRS 1915+105 does not have a regular hard state as in case of other black hole X-ray binaries, we distinguished soft and hard state by the spectral class ($\Phi$ for soft and $\chi$ for hard) as reported in \cite{2009Natur.458..481N}. We limit our analysis to only two observations since the motive of the study is to understand the change in state with respect to the change in wind, for which these two observations form an adequate sample. The two observations are comparatively less affected by pile up and provide a view of the source in two different spectral states. The observation performed on 2007-08-14 with an exposure of 49 ks (obsid:7485) corresponds to a soft state and the one performed on 2000-04-24 with an exposure of 30.3 ks (obsid:660) corresponds to a hard state. Other than being in two spectral states, the observations also differ in the strength and number of emission and absorption lines present in the spectra \citep{2009Natur.458..481N}. Both observations were in "pointing" mode and "timed" readout mode. We collected the archival data of these observations from the Chandra Data Archive (CDA) and used the CIAO 4.11 software package for the data reduction and analysis \citep{ciao_2006SPIE.6270E..1VF}. We reprocessed the data using "chandra\_repo" script to ensure data reduction using the latest Calibration data base (CALDB 4.8.2). In bright sources, the zeroth order position cannot be rightly identified by 'tgdetect' due to the hole in the zeroth order. Hence the zeroth order was identified by 'tg\_findzo' as suggested by 'tgdetect2'. For obsid:660, ACIS S-2 and ACIS S-3 were used to extract the signal, while for obsid:7485 ACIS S-1, ACIS S-2, ACIS S-3 and ACIS S-4 were used. We used only the HEG spectrum, as it has two times better energy resolution than the MEG at higher energies. Moreover, we were not interested in the energy range below 1.5 keV as GRS 1915+105 is heavily absorbed in that range and lower energies are more affected by pileup. We examined the +1 and -1 grating spectra separately to check for correspondence before combining them using "add\_grating\_orders". Finally, we used "mkgrmf" and "mkgarf" to generate the combined rmf and arf response files.  

\section{Light curve}
\begin{figure}
\centering
\resizebox{\hsize}{!}{\includegraphics[width=\hsize]{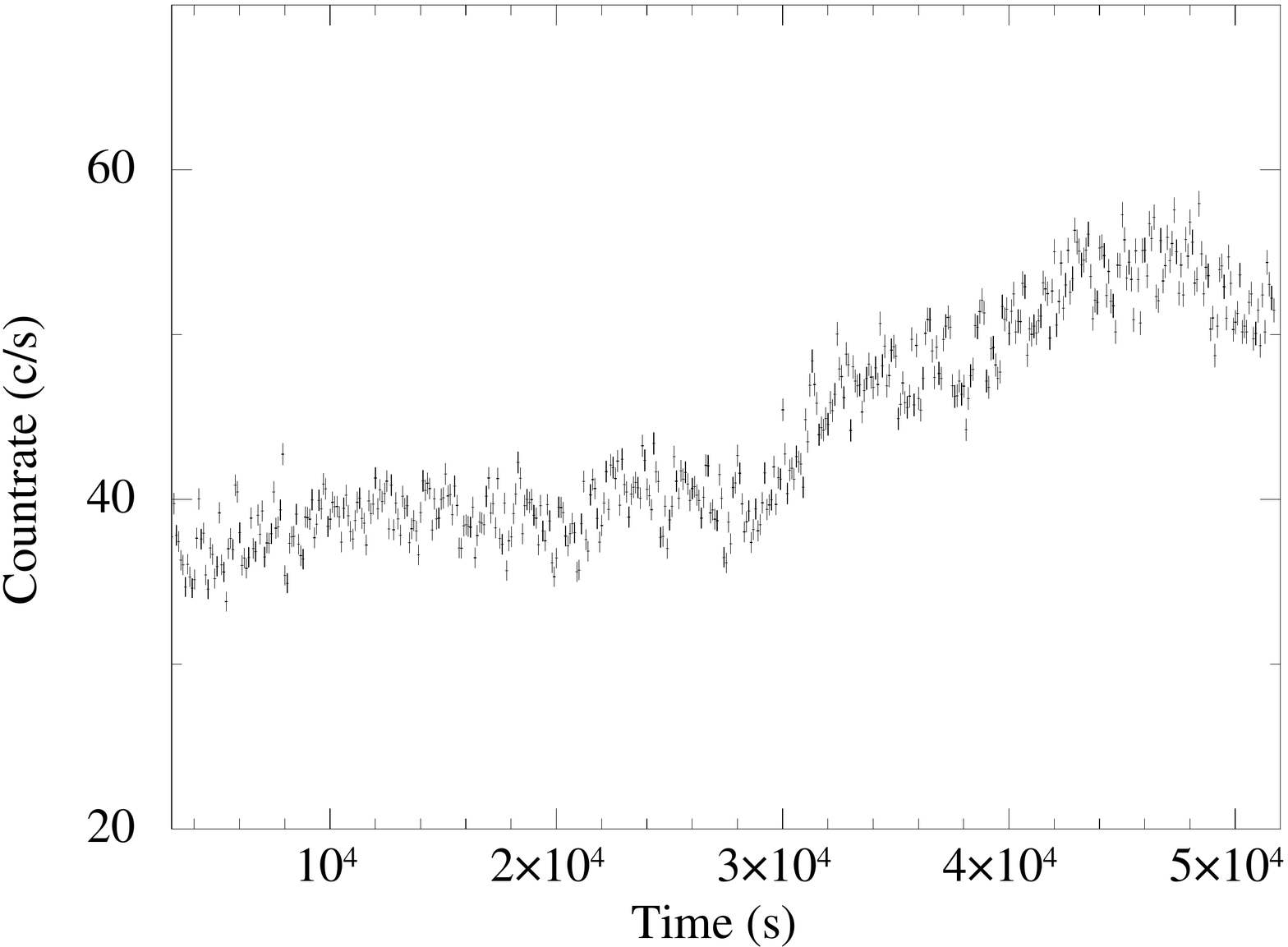}}
\caption{Light curve of GRS1915+105 in the soft state (obsid:7485) with a bin size of 100s.}
\label{7485_lc}
\end{figure}

\begin{figure}
\centering
\resizebox{\hsize}{!}{\includegraphics[width=\hsize]{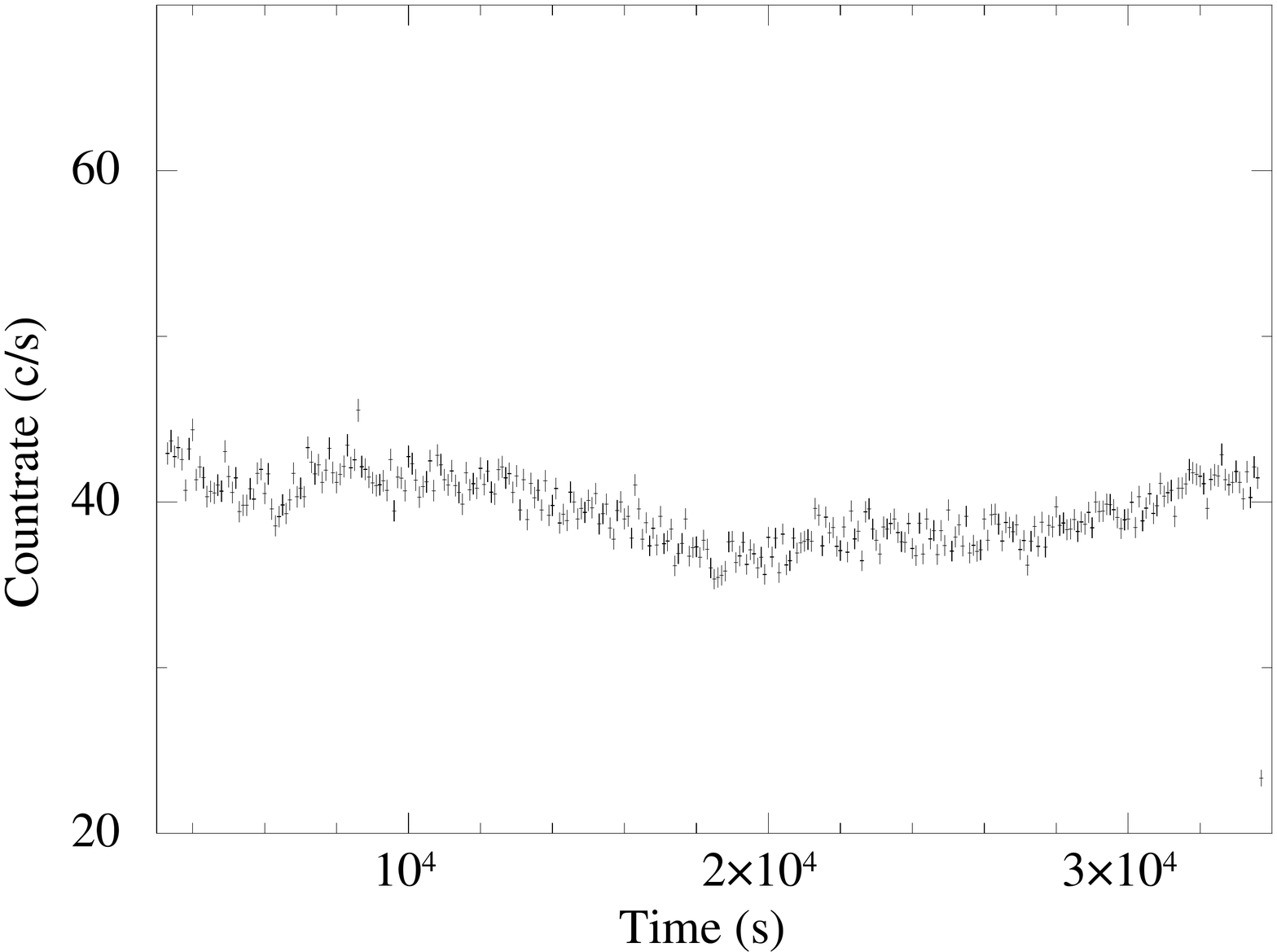}}
\caption{Light curve of GRS1915+105 in the hard state  with a bin size of 100s (obsid:660).}
\label{660_lc}
\end{figure}

The light curves for both observations were extracted using the "dmextract" tool. We considered the energy range from 1.5 keV to 7.2 keV. The light curves were analysed and re-binned using "lcurve" of XRONOS 5.22. Fig \ref{7485_lc} and Fig \ref{660_lc} show the light curves for both the soft and hard states, respectively. Based on the light curve and colour-colour diagrams, \cite{2000A&A...355..271B} classified GRS 1915+105 into 12 classes. \cite{2000A&A...355..271B} found that all the classes can be reduced to three fundamental states (A, B and C). In state B the accretion disc extends all the way down to innermost stable circular orbit (ISCO), and in state C the inner accretion disc is not visible due to reasons which are still debated. State A corresponds to a soft state with higher soft colour and higher countrate. In obsid:7485, there was an increase in flux observed during the observation. During the initial phase of this observation the source was in class $\phi$ (only state A) and in the later phase, the flux of the source increased to that of $\delta$ (transition between state A and state B). However, there is no change in the colour-colour diagram indicating it to be in bright $\phi$ class \citep{Ueda2009_2009ApJ...695..888U}. During obsid:660 the source was in $\chi$ class (only state C). 
\section{Spectral analysis}
The combined grating spectra of the 1st order were analysed using XSPEC 12.10.1. We used only HEG as the energy interval of interest is in the harder X-rays. 
We find an excess in soft X-rays below 2 keV in both the soft and hard state observations peaking at $\sim$ 0.12 keV for which the origin might not be physical and might be an instrumental feature. A similar feature for this source is also reported in XMM Newton observation \cite{Martocchia_2006A&A...448..677M}. We fitted the continuum spectra with a disc black body for the disc emission \citep[diskbb:][]{diskbb1_1984PASJ...36..741M,diskbb2_1986ApJ...308..635M}, a black body for the soft excess (bbody) and a power-law continuum for the X-ray corona  \citep[nthcomp:][]{nthcomp1_1996MNRAS.283..193Z,nthcomp2_1999MNRAS.309..561Z}, along with a neutral galactic absorption model \citep[tbabs:][]{TBABS_2000ApJ...542..914W} . However, since the spectra are piled up due to the high flux of GRS 1915+105, the parameters of the continuum model should be considered only as a phenomenological best-fit. Absorption and emission lines were modelled with a Gaussian ($zgauss$). Therefore, we concentrate our spectral analysis on the line properties, considering a phenomenological model for the continuum. The $\chi^2$ statistics was used for the fit and the errors are calculated at $1\sigma$ level. The modelled continuum parameters of all the observations are given in the Table \ref{fig_continumm_all}.
\begin{table*}
\caption{Continuum parameters obtained for the spectral fit for the soft (obsID:7485) and hard (obsID:660) state Chandra observations of GRS 1915+105.}
\label{fig_continumm_all}
\centering   
\renewcommand{\arraystretch}{1.5}
\begin{tabular}{c c c c c} 
\hline\hline  
Component & Parameter & Unit & soft & hard \\
\hline
TBabs    & N$_H$       & $\times10^{22}$~cm$^{-2}$ 		& $6.60^{+0.06}_{-0.06}$ &  $5.225^{+0.055}_{-0.037}$ \\
diskbb   & $T_{in}$ & keV & $2.28^{+0.03}_{-0.03}$  & $0.77^{+0.02}_{-0.01}$ \\
diskbb   & norm &  & $30.6^{+1.6}_{-6.4}$  & $1575^{+141}_{-324}$    \\
bbody   & $T$ & eV & $142.1^{+3.3}_{-3.3}$  & $126.3^{+7.7}_{-9.2}$   \\
bbody   & norm  &  & $4.82^{+1.01}_{-0.76}$  & $1.11^{+1.18}_{-0.41}$    \\
nthcomp & $\Gamma$ & & $>3.91$ &  $1.00^f$    \\
nthcomp & $KT_{e}$ &  keV  & $<1.00$ &  $2.17^f$\\
nthcomp & $KT_{bb}$ & keV  & $0.515^{+0.009}_{-0.002}$ &  $0.85^{f}$    \\
nthcomp & norm   & ph~keV$^{-1}$~cm$^{-2}$~s$^{-1}$  & $7.5^{+0.4}_{-0.3}$  & $0.04^{f}$    \\
flux  & & $\times 10^{-9}$ ergs cm$^{-2}$ s$^{-1}$ & $7.7^{+3.7}_{-0.9}$  & $6.7^{+1.2}_{-3.0}$ \\
$\chi^2$/dof & & & $2833/2635$  & $2766/2635$ \\
\hline
\end{tabular}
\end{table*}
\subsection{Soft state}
\begin{figure}[h]
\centering
\resizebox{\hsize}{!}{\includegraphics[width=1.\hsize]{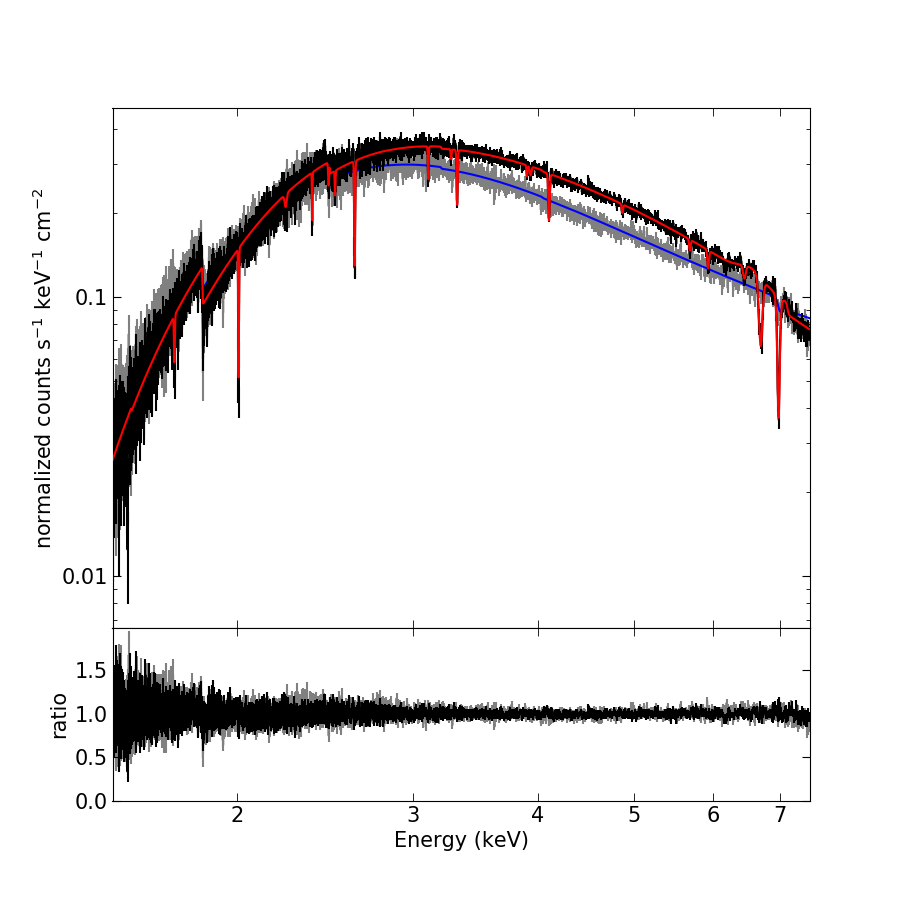}}
\caption{Top panel shows the spectrum of GRS 1915+105 in the soft state (obsid: 7485) and hard state (obsid: 660) after modelling the absorption and emission lines using the phenomenological Gaussian line profiles. The data in soft and hard states are plotted in black and grey, while the respective models in red and blue. The bottom panel shows the ratio of data and model.}
\label{Fig_7485_660_total}
\end{figure}

\begin{figure}[h]
\centering
\resizebox{\hsize}{!}{\includegraphics[width=1.\hsize]{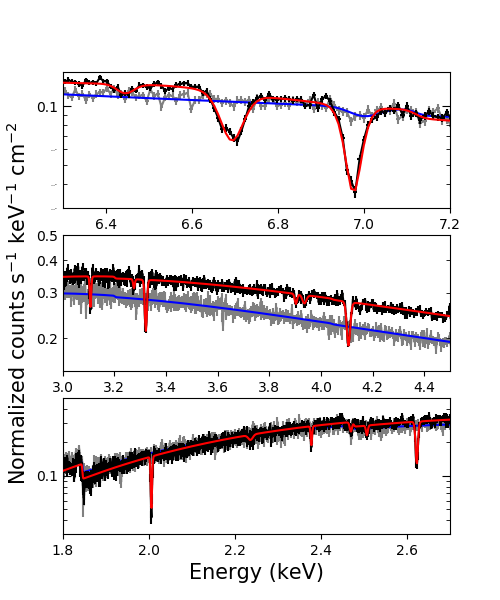}}
\caption{Zoomed spectrum (Top: 6.3-7.2 keV, middle: 3.0-4.5 keV, bottom: 1.8-2.7 keV) for the soft state (data in black and model in red) and hard state (data in grey and model in blue).}
\label{Fig_7485_660_total_zoomed}
\end{figure}

During obsid:7485, GRS 1915+105 was in class $\phi$, indicating a relatively soft spectrum dominated by the disc emission. Fig \ref{Fig_7485_660_total} shows the spectrum during the entire observation of 7485 and Fig \ref{Fig_7485_660_total_zoomed} shows the zoomed in version of the same at different energy ranges. We clearly find 20 absorption lines and 1 emission line in the spectrum. The lines were selected based on a minimum threshold of $\Delta$ $\chi^2$ of $15$. We fitted the absorption lines with inverted Gaussians and the emission line with a Gaussian. The line at 6.69 keV has previously been shown to split into two in the third order spectra \citep{Miller2016_2016ApJ...821L...9M}. Considering the above case, a double Gaussian was used to model the Fe XXV and Fe XXIV features, as these lines are close enough to overlap into a single observed line within the energy resolution of Chandra HETG 1st order. The energies of both the Fe XXV and Fe XXIV lines were tied to each other by an energy shift of $19.2$ eV, and leaving the norm and sigma free to vary. The $\chi {^2}$ value of the best fit is $2928.24$ for 2581 degrees of freedom (dof). There is a wide range of ionic species detected in the spectrum, the dominant being Al, Fe, Si, S, Ar, Ca, Cr, Mn, and Fe. The parameters of the detected lines are given in Table \ref{table_line_7485_epochall}.
\begin{table*}
\caption{ Line parameters for the soft state observation (obsid:7485).}
\label{table_line_7485_epochall}
\centering
\renewcommand{\arraystretch}{1.5}
\begin{tabular}{c c c c c c c} 
\hline\hline
Energy  & Sigma  & Norm  & Eqw & Line ID & Line Energy  & $\Delta$ $\chi^2$ \\ 
(eV) & (eV) & ($\times 10^{-3}$) & (eV) &  & & \\ 
\hline
&&&Absorption lines & & & \\ 
\hline
$1730.5^{+0.3}_{-0.4}$ & $1.66^{+0.44}_{-0.2}$ & $-7.9^{+1.1}_{-0.8}$ & $-1.69^{+0.24}_{-0.23}$ & $Al\ XIII\ 2p$  & $1.72769$ ($^2P_{1/2}$), $1.72899$ ( $^2P_{3/2}$ )  & $68$ \\
$2006.4^{+0.1}_{-0.1}$ & $1.01^{+0.06}_{-0.12}$ & $-8.0^{+0.2}_{-0.4}$ & $-3.25^{+0.2}_{-0.14}$ & $Si\ XIV\ 2p$ & $2.00433$ ($^2P_{1/2}$), $2.00608$ ($^2P_{3/2}$) & $1152$ \\
$2236.7^{+1.7}_{-1.8}$ & $5.03^{+1.44}_{-1.28}$ & $-2.8^{+0.5}_{-0.9}$ & $-1.42^{+0.44}_{-0.44}$ & NA & NA & $21$ \\ 
$2378.5^{+0.2}_{-1.8}$ & $\equiv 0$ & $-2.5^{+0.2}_{-0.3}$ & $-1.5^{+0.16}_{-0.17}$ & $Si\ XIV\ 3p$ & $2.37611$ ($^2P_{1/2}$), $2.37663$ ($^2P_{3/2}$)  & $99$ \\ 
$2470.7^{+0.7}_{-0.4}$ & $1.96^{+0.63}_{-0.58}$ & $-2.0^{+0.3}_{-0.3}$ & $-1.38^{+0.25}_{-0.21}$ & $S\ II\ 3p^d$& $2.4694^d$  & $46$ \\ 
$2506.9^{+0.4}_{-0.6}$ & $1.08^{+0.98}_{-0.88}$ & $-1.6^{+0.2}_{-0.3}$ & $-1.08^{+0.21}_{-0.19}$ & $Si\ XIV\ 4p$ & $2.50616$ ($^2P_{1/2}$), $2.50638$ ($^2P_{3/2}$)  & $44$\\
$2622.8^{+0.2}_{-0.2}$ & $2.19^{+0.14}_{-0.17}$ & $-6.2^{+0.2}_{-0.2}$ & $-4.76^{+0.26}_{-0.19}$ & $S\ XVI\ 2p$ & $2.61970$ ($^2P_{1/2}$), $2.62270$ ($^2P_{3/2}$)  & $977$ \\ 
$3108.8^{+0.3}_{-1.0}$ & $\equiv 0$ & $-1.5^{+0.1}_{-0.1}$ & $-1.8^{+0.1}_{-0.15}$ & $S\ XVI\ 3p$ & $3.10586$ ($^2P_{1/2}$), $3.10675$ ($^2P_{3/2}$)  & $176$ \\ 
$3277.0^{+0.8}_{-1.3}$ & $0.33^{+1.55}_{-0.33}$ & $-0.5^{+0.1}_{-0.1}$ & $-0.65^{+0.17}_{-0.19}$ & $S\ XVI\ 4p$ & $3.27589$ ($^2P_{1/2}$), $3.27627$ ($^2P_{3/2}$)  & $17$\\ 
$3322.7^{+0.2}_{-0.3}$ & $2.59^{+0.24}_{-0.39}$ & $-2.9^{+0.1}_{-0.1}$ & $-4.05^{+0.21}_{-0.2}$ & $Ar\ XVIII\ 2p$ & $3.31818$ ($^2P_{1/2}$), $3.32299$ ($^2P_{3/2}$) & $651$ \\ 
$3902.2^{+5.8}_{-0.6}$ & $\equiv 0$ & $-0.5^{+0.1}_{-0.1}$ & $-1.03^{+0.2}_{-0.27}$ & $Ca\ XIX\ 2p$ & $3.90226$ ($^1P_1$)  & $31$ \\ 
$3936.3^{+2.4}_{-1.3}$ & $5.96^{+1.92}_{-3.88}$ & $-0.7^{+0.1}_{-0.1}$ & $-1.47^{+0.33}_{-0.25}$ & $Ar\ XVIII\ 3p$ & $3.93429$ ($^2P_{1/2}$), $3.93572$ ($^2P_{3/2}$)  & $40$ \\ 
$4107.1^{+0.3}_{-0.4}$ & $4.69^{+0.44}_{-0.41}$ & $-2.5^{+0.1}_{-0.1}$ & $-5.91^{+0.23}_{-0.25}$ & $Ca\ XX\ 2p$ & $4.10015$ ($^2P_{1/2}$), $4.10750$ ($^2P_{3/2}$)  & $784$ \\ 
$4865.3^{+1.5}_{-2.8}$ & $1.02^{+3.08}_{-1.02}$ & $-0.4^{+0.1}_{-0.1}$ & $-1.34^{+0.27}_{-0.35}$ & $Ca\ XX\ 3p$ & $4.86192$ ($^2P_{1/2}$), $4.86410$ ($^2P_{3/2}$)  & $25$ \\ 
$5685.4^{+1.3}_{-4.8}$ & $\equiv 0$ & $-0.4^{+0.1}_{-0.1}$ & $-2.34^{+0.55}_{-0.39}$ & $Cr\ XXIII\ 2p$ & $5.68205$ ($^1P_{1}$)  & $29$ \\ 
$5931.3^{+0.8}_{-4.6}$ & $1.14^{+3.85}_{-1.14}$ & $-0.7^{+0.1}_{-0.1}$ & $-4.23^{+0.48}_{-0.41}$ & $Cr\ XXIV\ 2p$ & $5.91650$ ($^2P_{1/2}$), $5.93185$ ($^2P_{3/2}$)  & $73$ \\ 
$6446.2^{+4.5}_{-3.4}$ & $6.65^{+2.52}_{-7.21}$ & $-0.7^{+0.1}_{-0.1}$ & $-6.05^{+1.34}_{-1.22}$ & $Mn\ XXV\ 2p$ & $6.42356$ ($^2P_{1/2}$), $6.44166$ ($^2P_{3/2}$)  & $54$ \\ 
$6682.6^{+1.0}_{-1.5}$ & $23.18^{+2.48}_{-1.6}$ & $-2.0^{+0.1}_{-0.1}$ & $-16.72^{+8.53}_{-8.7}$ & $Fe\ XXIV\ 2p$  & $6.67644$ ($^2P_{1/2}$), $6.67915$ ($^2P_{3/2}$)   & $307$  \\ 
$6701.8^{+1.0}_{-1.5}$ & $23.12^{+1.42}_{-2.49}$ & $-2.4^{+0.1}_{-0.2}$ & $-19.66^{+7.12}_{-9.84}$ & $Fe\ XXV\ 2p$ & $6.6676$ ($^3P_{1}$), $6.7004$ ($^1P_{1}$)   & $429$ \\ 
$6975.6^{+0.8}_{-0.8}$ & $16.31^{+0.82}_{-0.82}$ & $-4.4^{+0.1}_{-0.1}$ & $-41.34^{+1.44}_{-1.0}$ & $Fe\ XXVI\ 2p$ & $6.95196$ ($^2P_{1/2}$), $6.97317$ ($^2P_{3/2}$)  & $2225$ \\ 
\hline
&&& Emission lines && &\\ 
\hline
$6538.9^{+19.6}_{-14.6}$ & $163.72^{+13.04}_{-14.8}$ & $4.1^{+0.3}_{-0.4}$ & $34.21^{+5.43}_{-5.66}$ & $Fe\ K$ &  & $182$ \\ 
\hline
\end{tabular}
\end{table*}

\subsection{Hard state}
During obsid:660, GRS 1915+105 was in class $\chi$, with a hard spectrum. The spectrum is rather featureless, with just a faint Fe XXVI absorption line.  The $\chi {^2}$ value for the best fit is $2916.25$ for 2636 dof. Fig \ref{Fig_7485_660_total} shows the spectrum and Table \ref{table_line_660} shows the line parameters. Further, a zoomed version of the hard state spectrum in different energy intervals is shown in Fig \ref{Fig_7485_660_total_zoomed}.

\begin{table*}
\caption{Line parameters in hard state observation (obsid:660).}
\label{table_line_660}
\centering 
\renewcommand{\arraystretch}{1.5}
\begin{tabular}{c c c c c c c} 
\hline\hline
Energy  & Sigma & Norm  & Eqw  &LineID & Line Energy (keV)\\ 
(eV) & (eV) & ($\times 10^{-3}$) & (eV) &  & \\
\hline
&&&Absorption lines \\
\hline
$6982.1^{+9.1}_{-10.1}$  & $13.17^{+9.14}_{-5.88}$ & $-0.6^{+0.1}_{-0.2}$ & $6.24^{+1.40}_{-2.50} $  & $Fe\ XXVI\ 2p$  & $6.95196$ ($^2P_{1/2}$), $6.97317$ ($^2P_{3/2}$) \\
\hline
\end{tabular}
\end{table*}



\section{Disc wind characterisation}
\subsection{Phenomenological scaling relations}
\begin{figure}[h]
\centering 
\resizebox{\hsize}{!}{\includegraphics[width=1.05\hsize]{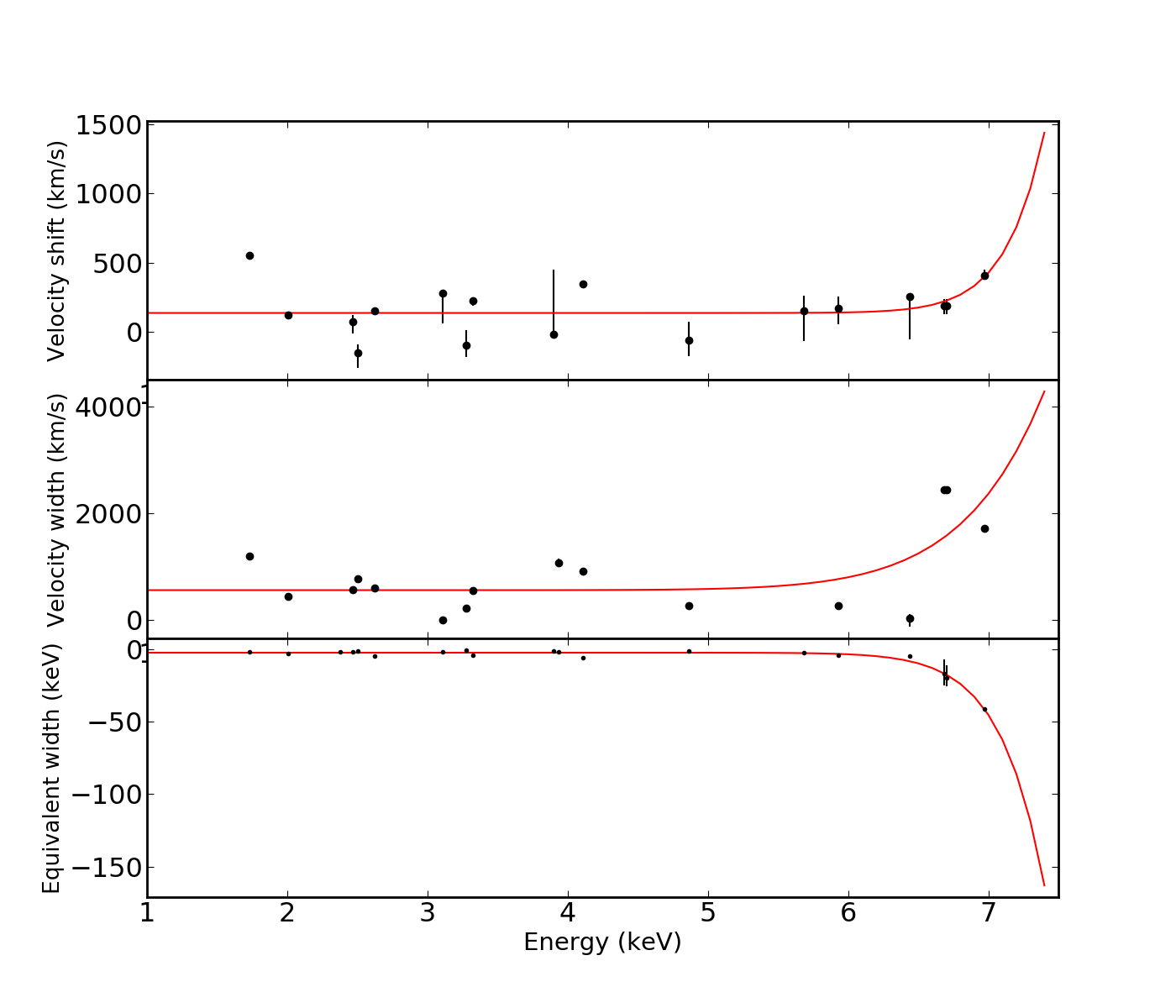}}
\caption{Velocity shift (top), velocity width (middle), and equivalent width (bottom) as a function of line energy for the soft state. In the top panel positive velocity shift corresponds to blue-shift.}
\label{fig_7485_vs_vw_eqw}
\end{figure}

From  Fig \ref{fig_7485_vs_vw_eqw} it  is  shown  that  the  profiles  of  the  velocity  shift, velocity width, and equivalent width of the absorption lines may be described by a power-law profile  ($A+B\times E^{\lambda}$) with respect to  the  ion  energies,  which  are  a  proxy  of  the  different  ionic species  and  ionisation  states. We took positive velocity shift as blue-shifted velocity. Within the uncertainties of the current high resolution X-ray instruments, it is not possible to decipher a clear non-linear trend if any. Upcoming high resolution instruments like XRISM and ATHENA will be able to decipher any non-linear trend \citep{XRISM_2020arXiv200304962X, ATHENA_2013arXiv1306.2307N}. The velocity shift is calculated with respect to the line energy calculated by the weighted average of different transitions with respect to their oscillator strength. The velocity width is calculated by V$_{w}$ = $2.355\sigma$c$/E_0$, where c is the speed of light, $E_0$ is the observed energy, V$_{w}$ is the velocity width, and $\sigma$ is the line width which is used to estimate the full width at half maximum (FWHM) of the line velocity profile. The parameters of the fit are listed in Table \ref{table_pl_fit}. The observed trends can be attributed to different wind ionisation states and different radii at which ions are present.
\\ \\ 
For a non linear profiles of the velocity shift, velocity width, and equivalent width, there can exist multiple components within the outflow. Most lines are consistent with a constant fit and some, particularly Fe XXVI and Fe XXV lines, show a large deviation. Either different components can be independent or the may be part of the same outflow spread over different radii and ionisation stages. Similar profiles in velocity shift and velocity width have also been seen in GRO J1655-40 by \cite{Kallman2009_2009ApJ...701..865K}. In GRO J1655-40 a large part of the outflow has similar velocity shift like in this case, with some ions significantly deviating from the constant, similar to the case shown here \citep{Kallman2009_2009ApJ...701..865K}. \cite{2017NatAs...1E..62F} showed that the non linear absorption line trend found in GRO J1655-40 was indicative of  different segments of a same magnetic outflow. In the next section we explore such a possibility for this source. 
\begin{table}
\caption{Parameters of the powerlaw fit on velocity shift, velocity width and equivalent width.}
\label{table_pl_fit}
\centering 
\renewcommand{\arraystretch}{1.5}
\begin{tabular}{c c c c} 
\hline\hline
Function  & A & B ( $\times10^{-21}$) & $\lambda$  \\ 
\hline
velocity shift & $134.72$ & $3.56$ $\times10^{-21}$ & $27.11$\\
velocity width & $554.88$ & $1.70$ $\times 10^{-08}$ & $16.58$ \\
eqw & $-2.26$ & $-3.22$ $\times 10^{-19}$ & $23.81$\\
\hline
\end{tabular}
\end{table}

\subsection{Magnetic disc wind modelling}
\begin{figure}[h]
\centering
\resizebox{\hsize}{!}{\includegraphics[width=1.\hsize]{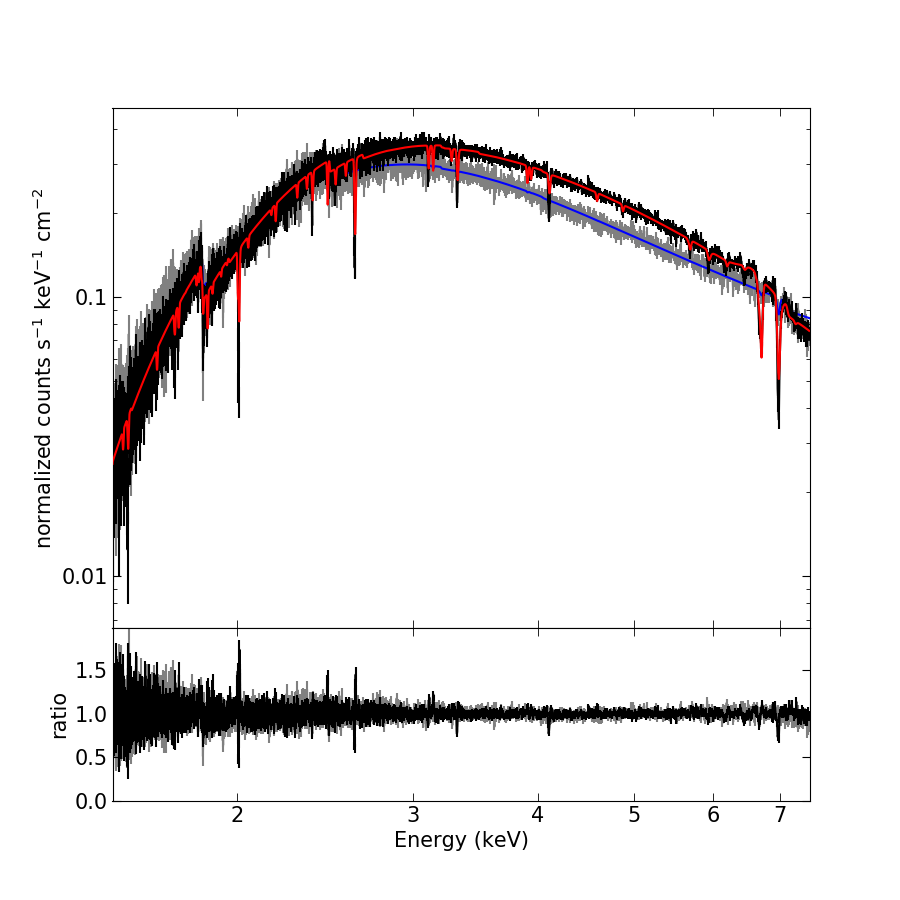}}
\caption{Top panel shows the spectrum of GRS 1915+105 in the soft state and hard state after modelling the absorption and emission lines with the MHD model. The data in soft and hard states are plotted in black and grey, while the respective models in red and blue. The bottom panel shows the ratio of data and model.}
\label{fig_7485_660_EpochALL_mhdmodel}
\end{figure}

\begin{figure}
\centering
\resizebox{\hsize}{!}{\includegraphics[width=1.\hsize]{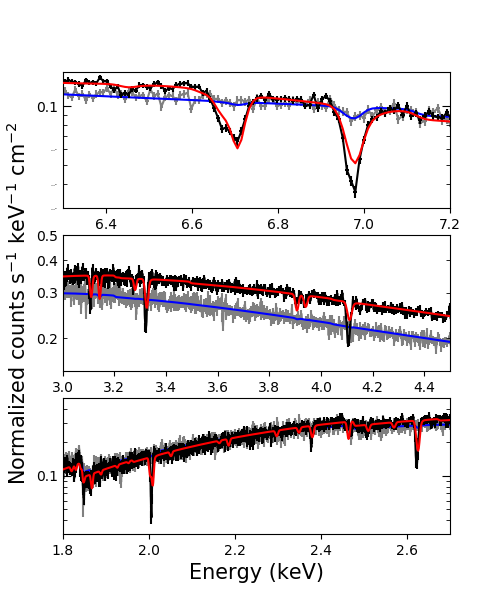}}
\caption{Zoomed (Top: 6.3-7.2 keV, middle: 3.0-4.5 keV, bottom: 1.8-2.7 keV) spectrum of GRS 1915+105 in the soft state (data in black and model in red) and hard state (data in grey and model in blue) after modelling with the MHD model.}
\label{fig_7485_660_mhdmodel_zoom}
\end{figure}

A magnetic origin of the wind has been suggested by several authors in both AGN and X-ray binaries \citep{Fukumura2010_2010ApJ...715..636F,2017NatAs...1E..62F,Blanford1982_1982MNRAS.199..883B,Contopoulos1994_1994ApJ...432..508C,Contopoulos1995_1995ApJ...450..616C,Ferreira1997_1997A&A...319..340F}. Plasma ejected and accelerated from the disc surface along the poloidal magnetic field lines can give rise to magnetically driven accretion disc winds \citep{2017NatAs...1E..62F}. This model is mass-invariant in a 2.5D, steady-state and axis-symmetric, wind structure for a set of wind parameters. It accounts for multiple absorption lines simultaneously in a self-consistent manner by modelling the internal structure of the wind, i.e. velocity, velocity gradient, density and ionization. This model can be used to determine the wind structure for a large range of parameters, such as distance, velocity and density, thus potentially constraining lines from a wide range of ionization states, both in the soft and hard X-ray band.\\ \\
The model is comprised of mainly two parts, the poloidal structure of the wind from a geometrically thin accretion disc and the ionisation state of the plasma computed by considering the local heating and cooling balance, and the ionisation equilibrium \citep{Fukumura2010_2010ApJ...715..636F,2017NatAs...1E..62F}. The ionising continuum is assumed to be a point source near the black hole. The model comprises of a single continuous wind structure in which different ions, at different ionisation states, represent parts of the same outflow. The higher the ionisation, column density and velocity, the closer the wind component is to the black hole. For a detailed description of the model we refer to \cite{2017NatAs...1E..62F,Fukumura2010_2010ApJ...715..636F}. The properties of the wind depend on the inclination angle ($\theta$), ionising spectral energy distribution (SED), and density normalisation ($n_{0}$) at the innermost launching radius which is assumed to be the innermost stable circular orbit (ISCO) for a Schwarzschild black hole. If the radii (r) and mass flux ($\Dot{m}$) are scaled by Schwarzschild radius ($R_S$) and Eddington rate ($\Dot{M}_{Edd}$) then the wind density profile can be expressed as $n(r)$ $=$ $n_{0}r^{-p}$. Then, the column density ($N_H$), the ionisation potential ($\xi$) and velocity ($v$) can be related to each other: $N_H$ $\propto$ $\xi^{(p-1)/(2-p)}$, $N_H$ $\propto$ $v^{2(p-1)}$, $v$ $\propto$ $\xi^{1/2(2-p)}$ \citep{2017NatAs...1E..62F}.
\\ \\
We calculated the detailed wind photo-ionisation structure using the ionising SED derived from extrapolating the observed Chandra unabsorbed continuum in the energy interval between E=13.6 eV and E=13.6 keV for both the soft and hard states.
However only the disc blackbody component and the power law component of the continuum was considered for the SED. We assumed that the wind starts at the ISCO in both the soft and the hard state. Since the inclination angle is well constrained for this source we fixed the inclination at $70.0$ degrees. Major line transitions and edges observable in the X-ray band have been implemented using a solar abundance apart from Fe, S, Si, where we left the abundances as parameter. We varied $p$ and $n_{0}$ in a self consistent way. For the fit $n_{0}$ was set to vary between $10^{15}$ $cm^{-3}$ and $3.2\times10^{18}$ $cm^{-3}$, and the density slope $p$ was set to vary between 0.9 and 1.5. For the soft state we got the best fit for $p$ = $1.378^{+0.001}_{-0.001}$ ($n$ $=$ $n_0(r/r_0)^{-1.38}$), and $n_{0}$ = $(19.8^{+0.6}_{-0.7})\times10^{17}$ cm$^{-3}$ with a $\chi^2$ of 4520 for 2631 dof (Fig \ref{fig_7485_660_EpochALL_mhdmodel}). For the hard state the best fit was obtained for $p$ = $1.378^{+0.005}_{-0.006}$ ($n$ $=$ $n_0(r/r_0)^{-1.18}$), and $n_{0}$ = $(0.50^{+0.05}_{-0.14})\times10^{17}$ cm$^{-3}$ with a $\chi^2$ of 2741 for 2631 dof. Additionally we also kept the abundances of Fe, S and Si as variable parameters in the fit. In the soft state we obtained a A$_{S}$ of $1.19^{+0.08}_{-0.07}$ and a lower limit of 2.92 for A$_{Fe}$ and an upper limit of 1.08 for A$_{Si}$ at 90 percent confidence. Similarly in the hard state we obtained a A$_{Fe}$ of $1.43^{+0.15}_{-0.11}$ and a lower limit of 2.42 and 2.0 for A$_{S}$ and A$_{Si}$ at 90 percent confidence. However, we note that the fit is not very sensitive to the varying abundances and consistent results are obtained also fixing the values to solar abundances. The broad-band modelling of the soft and hard states using the MHD wind model is shown in Fig \ref{fig_7485_660_EpochALL_mhdmodel}. The zoomed plot focused on the Fe XXV and Fe XXVI lines is shown in \ref{fig_7485_660_mhdmodel_zoom}.
[h]
\section{Discussion and conclusion}
\begin{figure}
\centering
\resizebox{\hsize}{!}{\includegraphics[width=1.0\hsize]{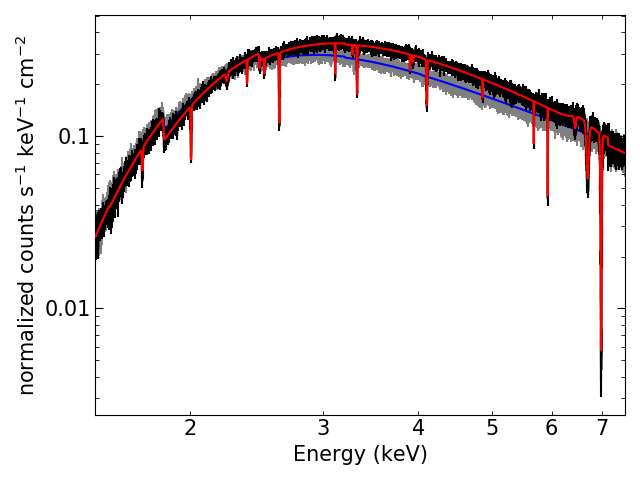}}
\caption{Simulated XRISM spectrum in the same energy band as Chandra HETGS for GRS 1915+105. Here we used the same Chandra HETG best-fit models for the spectra in soft (black) and hard (gray) states. The red and blue lines indicate the fitted phenomenological model.}
\label{XRISM_total}
\end{figure}

\begin{figure}
\centering
\resizebox{\hsize}{!}{\includegraphics[width=1.0\hsize]{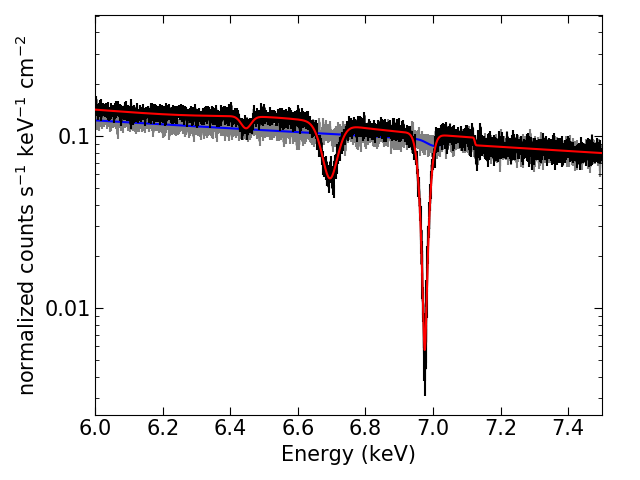}}
\caption{XRISM spectra of GRS 1915+105 zoomed in the 6.0 to 7.5 keV energy band. The soft and hard states are indicated in black and gray, respectively. The red and blue lines indicate the fitted phenomenological model.}
\label{XRISM_Fe}
\end{figure}

In this paper, we investigate the origin of the observed abrupt differences in the wind absorption line properties in a soft and hard state of GRS 1915+105 observed with Chandra HETG. We model the continuum spectra in both states using a black body and power law emission. We note that this is a phenomenological continuum model and we do not investigate in detail the continuum physical parameters because the main objective of this study is to investigate the narrow absorption features. As a second step, we fit the rich spectrum in the soft state with twenty absorption lines and one emission line using phenomenological Gaussian profiles. A faint Fe XXVI line found in the hard state was also fit with an inverted Gaussian profile.
\\ \\
We find phenomenological scaling relations of the velocity shift, velocity width and equivalent width of the absorption lines might follow a non linear pattern with respect to energy as a tracer of their ionisation, and thus suggesting multiple components in the wind. To explore this scenario further we fitted both the soft and hard state spectra using the MHD model that was used for the Chandra HETG spectrum of GRO J1655-40 in \cite{2017NatAs...1E..62F}. We find that in the soft state the best fit is obtained with a high wind density normalisation ($n_{0}$ = $(19.8^{+0.6}_{-0.7})\times10^{17}$ cm$^{-3}$) and for a radial wind density profile ($n$ $\propto$ $r^{-p}$) with $p \simeq 1.38$. Instead, in the hard state the best-fit values indicate a wind with a low density normalisation ($n_{0}$ = ($0.50^{+0.05}_{-0.14})\times10^{17}$ cm$^{-3}$), and a density profile with $p \simeq 1.38$. The main difference in the two absorption spectra is not dominated by the continuum change but instead it is due to an increase of two orders of magnitude in the wind density going from the hard to the soft state. Our model therefore predicts a persistent presence of disc winds even during hard state, although their spectroscopic appearance is much weaker. This result shows that the intrinsic wind condition changes internally in different states of this system, as previously speculated in other sources \citep{Neilsen2012_2012ApJ...750...27N, Ponti_2015MNRAS.446.1536P}. This also supports the suggested magnetic origin of the disc wind in GRS 1915+105 \citep{Miller2015_2015ApJ...814...87M, Miller2016_2016ApJ...821L...9M}. \\ \\
The wind density should in principle be a function of various disk micro physics and magnetic field properties, which are beyond the current state of our model. \cite{Fukumura_2014ApJ...780..120F} has investigated various types of morphology of MHD winds depending on the plasma property (e.g. the magnetic flux and plasma angular momentum). In this work we assume that the wind morphology remains almost unchanged throughout, considering that the global magnetic field pressure is dominant over radiation pressure in the framework of a generic MHD wind. In this case, the global field lines act like solid rigid wires \citep[e.g. ][]{Blanford1982_1982MNRAS.199..883B} being only weakly sensitive to mass accretion rate. It is conceivable, however, that local weak magnetic fields inside the disk, responsible for MRI (magneto-rotational instability), are more intimately coupled to the internal disk structure, and the mass accretion rate and accretion mode may well be a function of those local magnetic fields, while the global magnetic fields are persistent \citep[see][]{Mishra_Begelman_2020MNRAS.492.1855M}. Therefore, while a large-scale magnetic field structure is essentially unchanged, the cause of the wind density change could be due to the change in the local magnetic fields within the disk and/or accretion rate. This model is a steady state calculation that reflects the state of the disk-wind system at a given time. The wind could readjust over their entire range in time scales equal to their viscous time at their outer edge ($\sim$10$^6$ R$_g$, where R$_g$ is the Schwarzschild radius) and much faster locally (r $<<$ 10$^6$ R$_g$) for reasons we currently have not full control in our calculations. These time scales are of order of a few days for a 10 solar mass black hole, not inconsistent with the timescale of state transitions in BHBs. \\ \\
In this model we considered that wind is launched from nearby the ISCO in both the states. So, the characteristics of the MHD wind does not seem to depend strongly on a putative truncation radius of the disc in the hard state. 
Our fit results suggest that the wind structure changes with respect to the possible state change, and that photo-ionisation alone cannot explain the disappearance of the wind in the hard state, in accordance with what reported by, for example, \cite{Neilsen2012_2012ApJ...750...27N,Ponti_2015MNRAS.446.1536P}. It is conceivable that the change in the wind density may be linked to a change in the accretion disc density, accretion rate and/or geometry, which is itself responsible for the state change. \cite{Chakravorty_2016AN....337..429C} claim that a radial power-law profile of accretion rate determines the profile of the outflow, and for a higher power of the profile, the MHD wind is closer to the black hole. Our results are somewhat consistent with their findings, as we assumed a wind starting from ISCO and we obtain a relatively steeper profile in the soft state. They further claim that a wind is not possible in the canonical hard states of BHBs, however the hard state of GRS 1915+105 is peculiar because it does not follow the hysteresis loop in HID as in the case of other LMXBs. 
\\ \\
In principle, in the hard state, the wind could also be explained by a thermal origin. However, with the resolution and sensitivity of Chandra HETG, it is not possible to clearly differentiate between these two cases. An upcoming high resolution micro-calorimeter like XRISM might be useful to shed light on these rather controversial issues in more details. Also in the soft state we can see that an additional weak component might be required to fit all the lines properly. This can be attributed as a second component and might be due to a thermal wind. The magnetic and thermal component in the soft state would be consistent with the scenario that other authors would have seen as a multiple component wind \citep{Miller2015_2015ApJ...814...87M,Miller2016_2016ApJ...821L...9M}. So, in that scenario, the magnetic wind might be more variable and the thermal wind could be varying only with respect to the observed continuum. This means that the strong magnetic wind in the soft state was either absent or has become so weak for a detection in the hard state with Chandra HETG, and the thermal wind was present in both the states. This scenario can be similar to the hybrid thermal and magnetic wind suggested by \cite{Neilsen2012_2012ApJ...750...27N} for GRO J1655-40 in the hard state. In the soft state of GRO J1655-40, the two component wind found using photo-ionisation modelling by several authors \citep{Kallman2009_2009ApJ...701..865K,Miller2006_2006Natur.441..953M, Neilsen2012_2012ApJ...750...27N}, can be a part of a single component MHD outflow as suggested by \cite{2017NatAs...1E..62F}. However, even in \cite{2017NatAs...1E..62F}, we see that it is possible that a weak additional component might be required for a better fit. Another aspect that we did not incorporate in the spectra is the re-emission from the wind, broadened due to the Keplerian rotation at the photo-ionisation radius as suggested by \cite{Miller2015_2015ApJ...814...87M,Miller2016_2016ApJ...821L...9M}. We would incorporate the re-emission in a future work. We find that considering solar abundances we can derive a very good representation of the wind absorption. However, we note that some authors considered the possibility of super solar abundances for some elements when applying more simplistic models \citep[e.g.][]{Ueda2009_2009ApJ...695..888U}. 
We further estimate the physical parameters of the wind corresponding to the peak distribution of Fe XXVI ions for our best fit MHD model. Since each ion of a given charge state is produced through photo-ionisation over an radially extended distance along the MHD wind, the resultant ionic column is distributed over a finite range of distances. In the following estimates we provide the peak value which refers to the maximum column density and we include the range of values over which a quantity exceeds 50 per cent of the peak value. In the soft state, our best fit returns a Fe XXVI peak outflow velocity of $v_{out} = 343^{+146}_{-29}$ km s$^{-1}$, an ionisation parameter of log($\xi$) $=$ $4.31^{+0.76}_{-0.55}$ erg~s$^{-1}$~cm, and an equivalent hydrogen column density of $N_H \simeq 1.3\times10^{22}$ cm$^{-2}$. We estimate distance of the Fe XXVI absorber from the X-ray source of log(R/cm) = $13.6^{+1.0}_{-1.2}$. This does not mean that the launching radius of the wind is at this value, but it indicates that the peak of the ionisation parameter corresponds to this radius. Indeed, the MHD model considers a wind launched from a wide range of radii on the disc starting from the ISCO. The absorber located at distances lower than this is physically present but it is simply unobservable because even iron is fully ionized. In the units of Schwarzschild radius (R$_g$) the peak radius of Fe XXVI corresponds to $R \simeq 1.1\times10^7$ R$_g$. At such large radius the possibility of a thermal wind cannot be ruled out. However, we note that the broad ionisation range seen in the absorption lines, interpreted as a multiple component outflow by many authors \citep{Miller2015_2015ApJ...814...87M,Miller2016_2016ApJ...821L...9M}, supports a magnetic origin. Indeed, our MHD wind model in which every single absorber is physically coupled to the same continuous wind, is able to provide a very good representation of the absorption structure of the wind requiring a single stratified MHD disc wind.\\ \\
A recent work on accretion disc winds in H 1743-32 shows that a thermal wind is preferred to a magnetic wind \citep{Tomaru_2020MNRAS.494.3413T}. However they consider only the Fe XXV-XXVI lines and might not able to characterise a vast amount of soft X-ray lines found in GRS 1915+105. Our MHD model is able to provide a good characterisation of all the absorption lines seen in GRS 1915+105, because it intrinsically considers a stratified wind across the accretion disc. The thermal model in  \cite{Tomaru_2020MNRAS.494.3413T} and in general thermal wind models, imply a velocity trend in which the velocity decreases for decreasing launching radius, to the point in which the inner wind is highly ionised and static. Our data on GRS 1915+105 soft state show the opposite, we have a stratified wind with outflow velocity and velocity width of the lines increasing for increasing ionisation, which is equivalent to say that the outflow velocity increases for decreasing launching radius. The stratification of the ionisation in the same observation is shown in \cite{Miller2015_2015ApJ...814...87M} as well. Instead, our hard state observation, in which we only observe a faint Fe XXVI, can not exclude the possibility of thermal driving. As we already suggested, it's also possible that we have an hybrid situation, with a combination of thermal and MHD wind, with one of the two mechanisms appearing more prominently in different states of the source. \\ \\
Following \cite{2017NatAs...1E..62F,Tombesi_2015Natur.519..436T}, the local mass outflow rate from the wind can be well estimated combining the equation $\Dot{M}_{out}^{local}$ = $4\pi m_{p} n(r) r^2 v_{out}$ and the definition of the ionisation parameter $\xi$ = $L_{X}/n(r)r^2$, as $\Dot{M}_{out}$ = $4\pi m_{p} L_X v_{out}/\xi$. Considering our best fit Fe XXVI peak parameters and for an ionizing luminosity in the soft state of 4.32$\times$10$^{38}$ ergs s$^{-1}$ we estimate $\Dot{M}_{out}$ $\simeq$ $2.5\times$10$^{-9}$ M$_\odot$ yr$^{-1}$. The accretion rate, defined as $\Dot{M}_{acc}$ = $L_{X}/\eta c^2$, for the source in the soft state is $\simeq$ 7.7$\times$10$^{-8}$ M$_\odot$ yr$^{-1}$ for a typical value of $\eta = 0.1$. The estimate of power output through the wind can be derived as $\Dot{E}$ = $(1/2)\Dot{M}_{out}v_{out}^2$. Integrating the power output throughout the disc, the estimated wind energy output from the system is of the order of one percent of the luminosity.
\\ \\
Future upcoming high energy resolution instruments like XRISM \citep{XRISM_2020arXiv200304962X} and Athena \citep{ATHENA_2013arXiv1306.2307N} will be able to investigate this further. Weaker absorbers especially in the hard X-ray band (even during hard state if indeed present) can be better probed with the micro-calorimeters in XRISM/Athena. XRISM is planned to be launched in 2022. Here, we test its capabilities to investigate the accretion disc winds in GRS 1915+105 or any similar sources. So we simulate a spectra using our best fit phenomenological models in both soft and hard states. We used the XSPEC "fakeit" command to generate the spectra using an equivalent XRISM calorimeter response with 5 eV energy resolution and the associated background. Fig \ref{XRISM_total} and Fig \ref{XRISM_Fe} show the broadband and Fe K zoomed XRISM spectra in both the soft and hard states for a short exposure of 10 ks. The accuracy in the line energy, width and equivalent width for the Fe XXIV-XXVI lines with XRISM will represent more than an order of magnitude improvement compared to Chandra HETG. The line significance for the Fe XXVI, Fe XXV, Fe XXIV lines in the soft state and Fe XXVI in the hard state will improve more than a factor of three from the Chandra HETGS spectra. It is to be noted that the exposure for Chandra HETG in the soft and hard state were 49 ks and 30.3 ks compared to the just 10 ks for the simulated XRISM data. Hence, XRISM will very likely enable a time resolved study of accretion winds in this and most stellar mass black hole binaries.\\ 

It is important to note that X-ray binaries would be among the best candidates for detecting X-ray polarisation emission from the disc, the corona and circumnuclear matter \citep{XpolDovciak_2008JPhCS.131a2004D,XpolJeremy_2009ApJ...701.1175S,XpolJeremy2_2010ApJ...712..908S,XpolKallman_2015ApJ...815...53K, XpolTaverna_2020MNRAS.493.4960T}. We tested if the absorption lines in the GRS 1915+105 Chandra spectrum can affect the polarisation in the continuum in the context of the upcoming X-ray polarimetry mission IXPE \citep{IXPE_2016SPIE.9905E..17W}. We used the IXPEOBSSIM simulator, which takes as an input the spectrum and the polarisation angle and polarisation degree as a function of energy. The polarisation degree and angle as a function of energy, for different geometry of the disc and corona \citep[for example see][]{XpolDovciak_2008JPhCS.131a2004D}. However, for simplicity we considered a constant polarisation angle of $30\degree$ and constant polarisation degree of $5\%$. For the spectrum we used our Chandra soft state phenomenological model as an input. In the case 1, we used only the continuum, while in case 2 we included the absorption lines as well. From this simple test, we can conclude that the presence of unmodeled wind absorption in the soft state of GRS 1915+105 should not hamper the detection of polarisation from the continuum emission. However, the situation may be different if scattering from such a wind would be included. As a rough qualitative estimate, considering an MHD wind density profile of $n$ $=$ $n_{0}(r/r_0)^{-1.5}$, the total integrated column density would be $N_{H} \simeq 2 n_{0} r_{0}$, where $r_{0}$ is of the order of the ISCO. Substituting the ISCO for a non-spinning black hole with the mass of GRS 1915+105 and the density estimated from our best-fits, we find that in the soft state the wind may have a column of up to $N_{H} \simeq 4.4\times$10$^{25}$ cm$^{-2}$ in the equatorial direction, while in the hard state it would have a much lower value of $N_{H} \simeq 1.1\times$10$^{24}$ cm$^{-2}$. Hence, the wind may be mostly fully ionised but Compton thick in the equatorial direction in the soft state. Moreover, we note that emission and reflection from the disc wind in X-ray binaries has been suggested \citep[e.g.][]{Miller2015_2015ApJ...814...87M}. These evidences raise the possibility that some polarised signatures from the wind may be observable with IXPE if the source is caught in the soft state. We defer to future work for a quantification of the effect of disc winds on the polarised signal in the X-rays.




\begin{acknowledgements}
      AR and FT thank J. Neilsen, G. Matt and F. Muleri for the useful comments and discussions. We also thank the anonymous referee for his suggestions in improving this work. 
\end{acknowledgements}

\bibliographystyle{aa}
\bibliography{grs1915windmodelling}

\begin{thebibliography}{54}
\expandafter\ifx\csname natexlab\endcsname\relax\def\natexlab#1{#1}\fi

\bibitem[{{Begelman} {et~al.}(1983){Begelman}, {McKee}, \&
  {Shields}}]{Begelman1983_1983ApJ...271...70B}
{Begelman}, M.~C., {McKee}, C.~F., \& {Shields}, G.~A. 1983, \apj, 271, 70

\bibitem[{{Belloni} {et~al.}(2000){Belloni}, {Klein-Wolt}, {M{\'e}ndez}, {van
  der Klis}, \& {van Paradijs}}]{2000A&A...355..271B}
{Belloni}, T., {Klein-Wolt}, M., {M{\'e}ndez}, M., {van der Klis}, M., \& {van
  Paradijs}, J. 2000, \aap, 355, 271

\bibitem[{{Blandford} \& {Payne}(1982)}]{Blanford1982_1982MNRAS.199..883B}
{Blandford}, R.~D. \& {Payne}, D.~G. 1982, \mnras, 199, 883

\bibitem[{{Castor} {et~al.}(1975){Castor}, {Abbott}, \&
  {Klein}}]{Castor1975_1975ApJ...195..157C}
{Castor}, J.~I., {Abbott}, D.~C., \& {Klein}, R.~I. 1975, \apj, 195, 157

\bibitem[{{Chakravorty} {et~al.}(2016){Chakravorty}, {Petrucci}, {Ferreira},
  {Henri}, {Belmont}, {Clavel}, {Corbel}, {Rodriguez}, {Coriat}, {Drappeau}, \&
  {Malzac}}]{Chakravorty_2016AN....337..429C}
{Chakravorty}, S., {Petrucci}, P.~O., {Ferreira}, J., {et~al.} 2016,
  Astronomische Nachrichten, 337, 429

\bibitem[{{Contopoulos}(1994)}]{Contopoulos1994_1994ApJ...432..508C}
{Contopoulos}, J. 1994, \apj, 432, 508

\bibitem[{{Contopoulos}(1995)}]{Contopoulos1995_1995ApJ...450..616C}
{Contopoulos}, J. 1995, \apj, 450, 616

\bibitem[{{D{\'\i}az Trigo} \& {Boirin}(2016)}]{Trigo2016_2016AN....337..368D}
{D{\'\i}az Trigo}, M. \& {Boirin}, L. 2016, Astronomische Nachrichten, 337, 368

\bibitem[{{Done} {et~al.}(2018){Done}, {Tomaru}, \&
  {Takahashi}}]{done2018_2018MNRAS.473..838D}
{Done}, C., {Tomaru}, R., \& {Takahashi}, T. 2018, \mnras, 473, 838

\bibitem[{{Dov{\v{c}}iak} {et~al.}(2008){Dov{\v{c}}iak}, {Goosmann}, {Karas},
  \& {Matt}}]{XpolDovciak_2008JPhCS.131a2004D}
{Dov{\v{c}}iak}, M., {Goosmann}, R.~W., {Karas}, V., \& {Matt}, G. 2008, in
  Journal of Physics Conference Series, Vol. 131, Journal of Physics Conference
  Series, 012004

\bibitem[{Fender {et~al.}(2004)Fender, Belloni, \& Gallo}]{Fender_2004}
Fender, R.~P., Belloni, T.~M., \& Gallo, E. 2004, Monthly Notices of the Royal
  Astronomical Society, 355, 1105–1118

\bibitem[{{Ferreira}(1997)}]{Ferreira1997_1997A&A...319..340F}
{Ferreira}, J. 1997, \aap, 319, 340

\bibitem[{{Fruscione} {et~al.}(2006){Fruscione}, {McDowell}, {Allen},
  {Brickhouse}, {Burke}, {Davis}, {Durham}, {Elvis}, {Galle}, {Harris},
  {Huenemoerder}, {Houck}, {Ishibashi}, {Karovska}, {Nicastro}, {Noble},
  {Nowak}, {Primini}, {Siemiginowska}, {Smith}, \&
  {Wise}}]{ciao_2006SPIE.6270E..1VF}
{Fruscione}, A., {McDowell}, J.~C., {Allen}, G.~E., {et~al.} 2006, in Society
  of Photo-Optical Instrumentation Engineers (SPIE) Conference Series, Vol.
  6270, Society of Photo-Optical Instrumentation Engineers (SPIE) Conference
  Series, ed. D.~R. {Silva} \& R.~E. {Doxsey}, 62701V

\bibitem[{{Fukumura} {et~al.}(2010){Fukumura}, {Kazanas}, {Contopoulos}, \&
  {Behar}}]{Fukumura2010_2010ApJ...715..636F}
{Fukumura}, K., {Kazanas}, D., {Contopoulos}, I., \& {Behar}, E. 2010, \apj,
  715, 636

\bibitem[{{Fukumura} {et~al.}(2017){Fukumura}, {Kazanas}, {Shrader}, {Behar},
  {Tombesi}, \& {Contopoulos}}]{2017NatAs...1E..62F}
{Fukumura}, K., {Kazanas}, D., {Shrader}, C., {et~al.} 2017, Nature Astronomy,
  1, 0062

\bibitem[{{Fukumura} {et~al.}(2014){Fukumura}, {Tombesi}, {Kazanas}, {Shrader},
  {Behar}, \& {Contopoulos}}]{Fukumura_2014ApJ...780..120F}
{Fukumura}, K., {Tombesi}, F., {Kazanas}, D., {et~al.} 2014, \apj, 780, 120

\bibitem[{{Kallman} {et~al.}(2015){Kallman}, {Dorodnitsyn}, \&
  {Blondin}}]{XpolKallman_2015ApJ...815...53K}
{Kallman}, T., {Dorodnitsyn}, A., \& {Blondin}, J. 2015, \apj, 815, 53

\bibitem[{{Kallman} {et~al.}(2009){Kallman}, {Bautista}, {Goriely}, {Mendoza},
  {Miller}, {Palmeri}, {Quinet}, \&
  {Raymond}}]{Kallman2009_2009ApJ...701..865K}
{Kallman}, T.~R., {Bautista}, M.~A., {Goriely}, S., {et~al.} 2009, \apj, 701,
  865

\bibitem[{{Lee} {et~al.}(2002){Lee}, {Reynolds}, {Remillard}, {Schulz},
  {Blackman}, \& {Fabian}}]{2002ApJ...567.1102L}
{Lee}, J.~C., {Reynolds}, C.~S., {Remillard}, R., {et~al.} 2002, \apj, 567,
  1102

\bibitem[{{Makishima} {et~al.}(1986){Makishima}, {Maejima}, {Mitsuda}, {Bradt},
  {Remillard}, {Tuohy}, {Hoshi}, \& {Nakagawa}}]{diskbb2_1986ApJ...308..635M}
{Makishima}, K., {Maejima}, Y., {Mitsuda}, K., {et~al.} 1986, \apj, 308, 635

\bibitem[{{Martocchia} {et~al.}(2006){Martocchia}, {Matt}, {Belloni}, {Feroci},
  {Karas}, \& {Ponti}}]{Martocchia_2006A&A...448..677M}
{Martocchia}, A., {Matt}, G., {Belloni}, T., {et~al.} 2006, \aap, 448, 677

\bibitem[{{Miller}(2006)}]{2006AN....327..997M}
{Miller}, J.~M. 2006, Astronomische Nachrichten, 327, 997

\bibitem[{{Miller} {et~al.}(2019){Miller}, {Balakrishnan}, {Reynolds},
  {Trueba}, {Zoghbi}, {Kaastra}, {Kallman}, \&
  {Proga}}]{MillerAtel2019_2019ATel12743....1M}
{Miller}, J.~M., {Balakrishnan}, M., {Reynolds}, M.~T., {et~al.} 2019, The
  Astronomer's Telegram, 12743, 1

\bibitem[{{Miller} {et~al.}(2015){Miller}, {Fabian}, {Kaastra}, {Kallman},
  {King}, {Proga}, {Raymond}, \& {Reynolds}}]{Miller2015_2015ApJ...814...87M}
{Miller}, J.~M., {Fabian}, A.~C., {Kaastra}, J., {et~al.} 2015, \apj, 814, 87

\bibitem[{{Miller} {et~al.}(2006{\natexlab{a}}){Miller}, {Raymond}, {Fabian},
  {Steeghs}, {Homan}, {Reynolds}, {van der Klis}, \&
  {Wijnands}}]{Miller2006_2006Natur.441..953M}
{Miller}, J.~M., {Raymond}, J., {Fabian}, A., {et~al.} 2006{\natexlab{a}},
  \nat, 441, 953

\bibitem[{{Miller} {et~al.}(2016){Miller}, {Raymond}, {Fabian}, {Gallo},
  {Kaastra}, {Kallman}, {King}, {Proga}, {Reynolds}, \&
  {Zoghbi}}]{Miller2016_2016ApJ...821L...9M}
{Miller}, J.~M., {Raymond}, J., {Fabian}, A.~C., {et~al.} 2016, \apjl, 821, L9

\bibitem[{{Miller} {et~al.}(2004){Miller}, {Raymond}, {Fabian}, {Homan},
  {Nowak}, {Wijnands}, {van der Klis}, {Belloni}, {Tomsick}, {Smith},
  {Charles}, \& {Lewin}}]{2004ApJ...601..450M}
{Miller}, J.~M., {Raymond}, J., {Fabian}, A.~C., {et~al.} 2004, \apj, 601, 450

\bibitem[{{Miller} {et~al.}(2006{\natexlab{b}}){Miller}, {Raymond}, {Homan},
  {Fabian}, {Steeghs}, {Wijnands}, {Rupen}, {Charles}, {van der Klis}, \&
  {Lewin}}]{2006ApJ...646..394M}
{Miller}, J.~M., {Raymond}, J., {Homan}, J., {et~al.} 2006{\natexlab{b}}, \apj,
  646, 394

\bibitem[{{Miller} {et~al.}(2008){Miller}, {Raymond}, {Reynolds}, {Fabian},
  {Kallman}, \& {Homan}}]{Miller2008_2008ApJ...680.1359M}
{Miller}, J.~M., {Raymond}, J., {Reynolds}, C.~S., {et~al.} 2008, \apj, 680,
  1359

\bibitem[{{Mishra} {et~al.}(2020){Mishra}, {Begelman}, {Armitage}, \&
  {Simon}}]{Mishra_Begelman_2020MNRAS.492.1855M}
{Mishra}, B., {Begelman}, M.~C., {Armitage}, P.~J., \& {Simon}, J.~B. 2020,
  \mnras, 492, 1855

\bibitem[{{Mitsuda} {et~al.}(1984){Mitsuda}, {Inoue}, {Koyama}, {Makishima},
  {Matsuoka}, {Ogawara}, {Shibazaki}, {Suzuki}, {Tanaka}, \&
  {Hirano}}]{diskbb1_1984PASJ...36..741M}
{Mitsuda}, K., {Inoue}, H., {Koyama}, K., {et~al.} 1984, \pasj, 36, 741

\bibitem[{{Nandra} {et~al.}(2013){Nandra}, {Barret}, {Barcons}, {Fabian}, {den
  Herder}, {Piro}, {Watson}, {Adami}, {Aird}, {Afonso}, {Alexander},
  {Argiroffi}, {Amati}, {Arnaud}, {Atteia}, {Audard}, {Badenes}, {Ballet},
  {Ballo}, {Bamba}, {Bhardwaj}, {Stefano Battistelli}, {Becker}, {De Becker},
  {Behar}, {Bianchi}, {Biffi}, {B{\^\i}rzan}, {Bocchino}, {Bogdanov}, {Boirin},
  {Boller}, {Borgani}, {Borm}, {Bouch{\'e}}, {Bourdin}, {Bower}, {Braito},
  {Branchini}, {Branduardi-Raymont}, {Bregman}, {Brenneman}, {Brightman},
  {Br{\"u}ggen}, {Buchner}, {Bulbul}, {Brusa}, {Bursa}, {Caccianiga},
  {Cackett}, {Campana}, {Cappelluti}, {Cappi}, {Carrera}, {Ceballos},
  {Christensen}, {Chu}, {Churazov}, {Clerc}, {Corbel}, {Corral}, {Comastri},
  {Costantini}, {Croston}, {Dadina}, {D'Ai}, {Decourchelle}, {Della Ceca},
  {Dennerl}, {Dolag}, {Done}, {Dovciak}, {Drake}, {Eckert}, {Edge}, {Ettori},
  {Ezoe}, {Feigelson}, {Fender}, {Feruglio}, {Finoguenov}, {Fiore}, {Galeazzi},
  {Gallagher}, {Gandhi}, {Gaspari}, {Gastaldello}, {Georgakakis},
  {Georgantopoulos}, {Gilfanov}, {Gitti}, {Gladstone}, {Goosmann}, {Gosset},
  {Grosso}, {Guedel}, {Guerrero}, {Haberl}, {Hardcastle}, {Heinz}, {Alonso
  Herrero}, {Herv{\'e}}, {Holmstrom}, {Iwasawa}, {Jonker}, {Kaastra}, {Kara},
  {Karas}, {Kastner}, {King}, {Kosenko}, {Koutroumpa}, {Kraft}, {Kreykenbohm},
  {Lallement}, {Lanzuisi}, {Lee}, {Lemoine-Goumard}, {Lobban}, {Lodato},
  {Lovisari}, {Lotti}, {McCharthy}, {McNamara}, {Maggio}, {Maiolino}, {De
  Marco}, {de Martino}, {Mateos}, {Matt}, {Maughan}, {Mazzotta}, {Mendez},
  {Merloni}, {Micela}, {Miceli}, {Mignani}, {Miller}, {Miniutti}, {Molendi},
  {Montez}, {Moretti}, {Motch}, {Naz{\'e}}, {Nevalainen}, {Nicastro}, {Nulsen},
  {Ohashi}, {O'Brien}, {Osborne}, {Oskinova}, {Pacaud}, {Paerels}, {Page},
  {Papadakis}, {Pareschi}, {Petre}, {Petrucci}, {Piconcelli}, {Pillitteri},
  {Pinto}, {de Plaa}, {Pointecouteau}, {Ponman}, {Ponti}, {Porquet}, {Pounds},
  {Pratt}, {Predehl}, {Proga}, {Psaltis}, {Rafferty}, {Ramos-Ceja}, {Ranalli},
  {Rasia}, {Rau}, {Rauw}, {Rea}, {Read}, {Reeves}, {Reiprich}, {Renaud},
  {Reynolds}, {Risaliti}, {Rodriguez}, {Rodriguez Hidalgo}, {Roncarelli},
  {Rosario}, {Rossetti}, {Rozanska}, {Rovilos}, {Salvaterra}, {Salvato}, {Di
  Salvo}, {Sanders}, {Sanz-Forcada}, {Schawinski}, {Schaye}, {Schwope},
  {Sciortino}, {Severgnini}, {Shankar}, {Sijacki}, {Sim}, {Schmid}, {Smith},
  {Steiner}, {Stelzer}, {Stewart}, {Strohmayer}, {Str{\"u}der}, {Sun}, {Takei},
  {Tatischeff}, {Tiengo}, {Tombesi}, {Trinchieri}, {Tsuru}, {Ud-Doula},
  {Ursino}, {Valencic}, {Vanzella}, {Vaughan}, {Vignali}, {Vink}, {Vito},
  {Volonteri}, {Wang}, {Webb}, {Willingale}, {Wilms}, {Wise}, {Worrall},
  {Young}, {Zampieri}, {In't Zand}, {Zane}, {Zezas}, {Zhang}, \&
  {Zhuravleva}}]{ATHENA_2013arXiv1306.2307N}
{Nandra}, K., {Barret}, D., {Barcons}, X., {et~al.} 2013, arXiv e-prints,
  arXiv:1306.2307

\bibitem[{{Neilsen}(2013)}]{Neilsen2013_2013AdSpR..52..732N}
{Neilsen}, J. 2013, Advances in Space Research, 52, 732

\bibitem[{{Neilsen} {et~al.}(2018){Neilsen}, {Cackett}, {Remillard}, {Homan},
  {Steiner}, {Gendreau}, {Arzoumanian}, {Prigozhin}, {LaMarr}, {Doty},
  {Eikenberry}, {Tombesi}, {Ludlam}, {Kara}, {Altamirano}, \&
  {Fabian}}]{Neilsen2018_2018ApJ...860L..19N}
{Neilsen}, J., {Cackett}, E., {Remillard}, R.~A., {et~al.} 2018, \apjl, 860,
  L19

\bibitem[{{Neilsen} \& {Homan}(2012)}]{Neilsen2012_2012ApJ...750...27N}
{Neilsen}, J. \& {Homan}, J. 2012, \apj, 750, 27

\bibitem[{{Neilsen} \& {Lee}(2009)}]{2009Natur.458..481N}
{Neilsen}, J. \& {Lee}, J.~C. 2009, \nat, 458, 481

\bibitem[{{Neilsen} {et~al.}(2016){Neilsen}, {Rahoui}, {Homan}, \&
  {Buxton}}]{Neilsen2016_2016ApJ...822...20N}
{Neilsen}, J., {Rahoui}, F., {Homan}, J., \& {Buxton}, M. 2016, \apj, 822, 20

\bibitem[{{Ponti} {et~al.}(2015){Ponti}, {Bianchi}, {Mu{\~n}oz-Darias}, {De
  Marco}, {Dwelly}, {Fender}, {Nandra}, {Rea}, {Mori}, {Haggard}, {Heinke},
  {Degenaar}, {Aramaki}, {Clavel}, {Goldwurm}, {Hailey}, {Israel}, {Morris},
  {Rushton}, \& {Terrier}}]{Ponti_2015MNRAS.446.1536P}
{Ponti}, G., {Bianchi}, S., {Mu{\~n}oz-Darias}, T., {et~al.} 2015, \mnras, 446,
  1536

\bibitem[{{Ponti} {et~al.}(2012){Ponti}, {Fender}, {Begelman}, {Dunn},
  {Neilsen}, \& {Coriat}}]{2012MNRAS.422L..11P}
{Ponti}, G., {Fender}, R.~P., {Begelman}, M.~C., {et~al.} 2012, \mnras, 422,
  L11

\bibitem[{{Schnittman} \& {Krolik}(2009)}]{XpolJeremy_2009ApJ...701.1175S}
{Schnittman}, J.~D. \& {Krolik}, J.~H. 2009, \apj, 701, 1175

\bibitem[{{Schnittman} \& {Krolik}(2010)}]{XpolJeremy2_2010ApJ...712..908S}
{Schnittman}, J.~D. \& {Krolik}, J.~H. 2010, \apj, 712, 908

\bibitem[{{Taverna} {et~al.}(2020){Taverna}, {Zhang}, {Dov{\v{c}}iak},
  {Bianchi}, {Bursa}, {Karas}, \& {Matt}}]{XpolTaverna_2020MNRAS.493.4960T}
{Taverna}, R., {Zhang}, W., {Dov{\v{c}}iak}, M., {et~al.} 2020, \mnras, 493,
  4960

\bibitem[{{Tomaru} {et~al.}(2020){Tomaru}, {Done}, {Ohsuga}, {Odaka}, \&
  {Takahashi}}]{Tomaru_2020MNRAS.494.3413T}
{Tomaru}, R., {Done}, C., {Ohsuga}, K., {Odaka}, H., \& {Takahashi}, T. 2020,
  \mnras, 494, 3413

\bibitem[{{Tombesi} {et~al.}(2010){Tombesi}, {Cappi}, {Reeves}, {Palumbo},
  {Yaqoob}, {Braito}, \& {Dadina}}]{Tombesi2010_2010A&A...521A..57T}
{Tombesi}, F., {Cappi}, M., {Reeves}, J.~N., {et~al.} 2010, \aap, 521, A57

\bibitem[{{Tombesi} {et~al.}(2015){Tombesi}, {Mel{\'e}ndez}, {Veilleux},
  {Reeves}, {Gonz{\'a}lez-Alfonso}, \&
  {Reynolds}}]{Tombesi_2015Natur.519..436T}
{Tombesi}, F., {Mel{\'e}ndez}, M., {Veilleux}, S., {et~al.} 2015, \nat, 519,
  436

\bibitem[{{Trueba} {et~al.}(2019){Trueba}, {Miller}, {Kaastra}, {Zoghbi},
  {Fabian}, {Kallman}, {Proga}, \& {Raymond}}]{Trueba2019_2019ApJ...886..104T}
{Trueba}, N., {Miller}, J.~M., {Kaastra}, J., {et~al.} 2019, \apj, 886, 104

\bibitem[{{Ueda} {et~al.}(2004){Ueda}, {Murakami}, {Yamaoka}, {Dotani}, \&
  {Ebisawa}}]{Ueda2004_2004ApJ...609..325U}
{Ueda}, Y., {Murakami}, H., {Yamaoka}, K., {Dotani}, T., \& {Ebisawa}, K. 2004,
  \apj, 609, 325

\bibitem[{{Ueda} {et~al.}(2009){Ueda}, {Yamaoka}, \&
  {Remillard}}]{Ueda2009_2009ApJ...695..888U}
{Ueda}, Y., {Yamaoka}, K., \& {Remillard}, R. 2009, \apj, 695, 888

\bibitem[{{Weisskopf} {et~al.}(2016){Weisskopf}, {Ramsey}, {O'Dell}, {Tennant},
  {Elsner}, {Soffitta}, {Bellazzini}, {Costa}, {Kolodziejczak}, {Kaspi},
  {Muleri}, {Marshall}, {Matt}, \& {Romani}}]{IXPE_2016SPIE.9905E..17W}
{Weisskopf}, M.~C., {Ramsey}, B., {O'Dell}, S., {et~al.} 2016, in Society of
  Photo-Optical Instrumentation Engineers (SPIE) Conference Series, Vol. 9905,
  \procspie, 990517

\bibitem[{{Wilms} {et~al.}(2000){Wilms}, {Allen}, \&
  {McCray}}]{TBABS_2000ApJ...542..914W}
{Wilms}, J., {Allen}, A., \& {McCray}, R. 2000, \apj, 542, 914

\bibitem[{{Woods} {et~al.}(1996){Woods}, {Klein}, {Castor}, {McKee}, \&
  {Bell}}]{woods1996_1996ApJ...461..767W}
{Woods}, D.~T., {Klein}, R.~I., {Castor}, J.~I., {McKee}, C.~F., \& {Bell},
  J.~B. 1996, \apj, 461, 767

\bibitem[{{XRISM Science Team}(2020)}]{XRISM_2020arXiv200304962X}
{XRISM Science Team}. 2020, arXiv e-prints, arXiv:2003.04962

\bibitem[{{Zdziarski} {et~al.}(1996){Zdziarski}, {Johnson}, \&
  {Magdziarz}}]{nthcomp1_1996MNRAS.283..193Z}
{Zdziarski}, A.~A., {Johnson}, W.~N., \& {Magdziarz}, P. 1996, \mnras, 283, 193

\bibitem[{{{\.Z}ycki} {et~al.}(1999){{\.Z}ycki}, {Done}, \&
  {Smith}}]{nthcomp2_1999MNRAS.309..561Z}
{{\.Z}ycki}, P.~T., {Done}, C., \& {Smith}, D.~A. 1999, \mnras, 309, 561

\end{thebibliography}
\begin{appendix}
\FloatBarrier
\section{Line properties in the soft state}
 Table \ref{velocity_shift_epoch_all} shows the velocity shift, velocity drift and equivalent width in the soft state observation (obsid:7485).

\begin{table}[h]
\caption{ Velocity shifts, velocity width and equivalent width in absorption lines during the whole (obsid:7485).}
\label{velocity_shift_epoch_all}
\renewcommand{\arraystretch}{1.5}
\begin{tabular}{c c c c} 
\hline\hline
\centering
Energy  & Velocity Shift  & Velocity Width & Eqw \\ 
(eV) & (km/s) & (km/s) & (eV) \\ \hline
$1730.5^{+0.3}_{-0.4}$ & $381.4^{+51.9}_{-69.3}$ & $677.7^{+13.6}_{-6.2}$ & $-1.69^{+0.24}_{-0.23}$ \\ 
$2006.4^{+0.1}_{-0.1}$ & $179.4^{+14.9}_{-14.9}$ & $355.6^{+1.7}_{-3.5}$ & $-3.25^{+0.2}_{-0.14}$ \\ 
$2378.5^{+0.2}_{-1.8}$ & $264.9^{+25.2}_{-226.8}$ & $\equiv 0$
& $-1.5^{+0.16}_{-0.17}$ \\ 
$2470.7^{+0.7}_{-0.4}$ & $157.9^{+85.0}_{-48.5}$ & $560.5^{+16.3}_{-15.0}$ & $-1.38^{+0.25}_{-0.21}$ \\ 
$2506.9^{+0.4}_{-0.6}$ & $71.8^{+47.9}_{-71.8}$ & $304.4^{+25.2}_{-22.7}$ & $-1.08^{+0.21}_{-0.19}$ \\ 
$2622.8^{+0.2}_{-0.2}$ & $183.0^{+22.9}_{-22.9}$ & $589.9^{+3.5}_{-4.3}$ & $-4.76^{+0.26}_{-0.19}$ \\ 
$3108.8^{+0.3}_{-1.0}$ & $241.3^{+28.9}_{-96.4}$ & $\equiv 0$
& $-1.8^{+0.1}_{-0.15}$ \\ 
$3277.0^{+0.8}_{-1.3}$ & $82.4^{+73.2}_{-119.0}$ & $71.1^{+34.9}_{-7.4}$ & $-0.65^{+0.17}_{-0.19}$ \\ 
$3322.7^{+0.2}_{-0.3}$ & $198.6^{+18.0}_{-27.1}$ & $550.7^{+5.4}_{-8.7}$ & $-4.05^{+0.21}_{-0.2}$ \\ 
$3902.2^{+5.8}_{-0.6}$ & $-7.7^{+445.9}_{-46.1}$ & $\equiv 0$
& $-1.03^{+0.2}_{-0.27}$ \\ 
$3936.3^{+2.4}_{-1.3}$ & $99.1^{+182.9}_{-99.0}$ & $1069.7^{+39.5}_{-79.8}$ & $-1.47^{+0.33}_{-0.25}$ \\ 
$4107.1^{+0.3}_{-0.4}$ & $241.0^{+21.9}_{-29.2}$ & $806.8^{+8.9}_{-8.3}$ & $-5.91^{+0.23}_{-0.25}$ \\ 
$4865.3^{+1.5}_{-2.8}$ & $141.8^{+92.4}_{-172.6}$ & $148.1^{+57.0}_{-18.9}$ & $-1.34^{+0.27}_{-0.35}$ \\ 
$5685.4^{+1.3}_{-4.8}$ & $174.1^{+68.6}_{-253.1}$ & $\equiv 0$
& $-2.34^{+0.55}_{-0.39}$ \\ 
$5931.3^{+0.8}_{-4.6}$ & $232.7^{+40.4}_{-232.5}$ & $135.8^{+64.5}_{-19.1}$ & $-4.23^{+0.48}_{-0.41}$ \\ 
$6446.2^{+4.5}_{-3.4}$ & $493.3^{+209.1}_{-158.0}$ & $728.8^{+40.5}_{-115.8}$ & $-6.05^{+1.34}_{-1.22}$ \\ 
$6682.6^{+1.0}_{-1.5}$ & $193.0^{+44.9}_{-67.3}$ & $2450.6^{+39.1}_{-25.2}$ & $-16.72^{+8.53}_{-8.7}$ \\ 
$6701.8^{+1.0}_{-1.5}$ & $192.5^{+44.7}_{-67.1}$ & $2437.3^{+22.4}_{-39.2}$ & $-19.66^{+7.12}_{-9.84}$ \\ 
$6975.6^{+0.8}_{-0.8}$ & $408.6^{+34.4}_{-34.4}$ & $1651.9^{+12.7}_{-12.7}$ & $-41.34^{+1.44}_{-1.0}$ \\ 
\hline
\end{tabular}
\end{table}

\end{appendix}

\end{document}